\documentclass[fleqn,12pt]{elsarticle}

\usepackage{graphicx,color}
\usepackage{amsmath,amssymb}
\newcommand{\Slash}[1]{{\ooalign{\hfil/\hfil\crcr$#1$}}}

\newcommand{\Nf}{N_{\rm f}}

\newcommand{\vp}{\vec{p}}

\newcommand{\la}{\langle}
\newcommand{\ra}{\rangle}

\newcommand{\calL}{\mathcal{L}}

\newcommand{\rmi}{\mathrm{i}}
\newcommand{\rme}{\mathrm{e}}

\journal{Physics Letter B}

\begin{document}

\begin{frontmatter}
\title{Pseudo Nambu-Goldstone modes in neutron stars}
\author {Toru Kojo}
\address {Key Laboratory of Quark and Lepton Physics (MOE) and Institute of Particle Physics, Central China Normal University, Wuhan 430079, China}
%
\begin{abstract}
If quarks and gluons are either gapped or confined in neutron stars (NSs), the most relevant light modes are Nambu-Goldstone (NG) modes. We study NG modes within a schematic quark model whose parameters {\it at high density} are constrained by the two-solar mass constraint. Our model has the color-flavor-locked phase at high density, with the effective couplings as strong as in hadron physics. We find that strong coupling effects make NG modes more massive than in weak coupling predictions, and would erase several phenomena caused by the stressed pairings in mismatched Fermi surfaces. For instance, we found that charged kaons, which are dominated by diquark and anti-diquark components, are not light enough to condense at strong coupling. Implications for gravitational wave signals for NS-NS mergers are also briefly discussed.
\end{abstract}
\begin{keyword} 
Dense QCD, effective models, neutron stars
\end{keyword}
\end{frontmatter}

\section{Introduction}
\label{sec:intro}

Neutron stars are unique cosmic laboratories to study cold dense QCD. The observations of two-solar mass ($2M_\odot$) neutron stars (NSs) \cite{2m_1,2m_2} tell us that equations of state at high baryon density, $n_B \gtrsim 5n_0$ ($n_0 \simeq 0.16\,{\rm fm}^{-3}$: nuclear saturation density), must be very stiff to prevent the star from gravitational collapsing. Meanwhile low density equations of state at $n_B \lesssim 2n_0$ should be softer than the previous thoughts, as suggested by studies of neutron star radii \cite{Steiner:2010fz,Ozel:2010fw,Ozel:2015fia,Guillot:2014lla,Heinke:2014xaa}, nuclear symmetry energy, heavy ion data \cite{Danielewicz:2002pu}, and predictions of chiral effective theories \cite{Epelbaum:2008ga} that are combined with sophisticated many-body calculations \cite{Gandolfi:2011xu,Gandolfi:2015jma}. Then the theoretical challenge is how to reconcile these tendencies at low and high density \cite{Kojo:2014rca,Kojo:2015fua}; to connect soft and stiff equations of state, there must be a region where the sound velocity $c_s = ( \partial P/\partial \varepsilon)^{1/2}$ is large, but this tends to violate the causality constraint $c_s \le 1$. Also, soft-to-stiff equations of state does not allow us to implement strong first order phase transitions which soften equations of state at high density\footnote{If we assume very stiff hadronic equations of state at low density, we may allow strong first order phase transitions from very stiff hadronic matter to quark matter stiff enough to pass the $2M_\odot$ constraint \cite{Benic:2014jia}.}. In this way the constraints from high and low density {\it together} limit the possible classes of equations of state, and from which one can extract useful insights into matter \cite{Lattimer:2006xb}.

While the current observations have already provided significant constraints, equations of state alone do not finalize our understanding of dense matter. It is desirable to study dynamic and thermal aspects of matter which originate from excitation modes whose properties are very sensitive to the phase structure. The predictions for thermal equations of state are very important for the gravitational wave astronomy \cite{Yunes:2016jcc,Baiotti:2016qnr}; the first direct detection of gravitational waves of a binary black-holes has been made in September 2015 \cite{Abbott:2016blz}, and the second detection three months later \cite{Abbott:2016nmj}. We also expect gravitational waves from NS-NS mergers in near future. It has been argued that the patterns of gravitational waves can discriminate different equations of state by comparing the observations with the waveforms predicted by numerical relativity with input equations of state \cite{Read:2013zra}.

The purpose of this paper is to study the low energy modes at high density, in a setup consistent with neutron star constraints at zero temperature. Our underlying physical picture is based on a 3-window description\footnote{Somewhat indirect, but less model-dependent approach is to interpolate the nuclear and perturbative quark matter equations of state. This has been carried out with perturbative calculations to three loop order \cite{Kurkela:2014vha}.} for dense QCD matter which has been studied in recent works \cite{Kojo:2014rca,Masuda:2012kf,Alvarez-Castillo:2013spa,Hell:2014xva}. The 3-window picture consists of:  purely nuclear matter for $n_B\lesssim 2n_0$; quark matter for $n_B \gtrsim 5n_0$; and crossover (or weak 1st order) picture\footnote{The crossover picture has been also discussed in an effective Lagrangian framework in \cite{Paeng:2015noa}, where matter changes its character but does not induce the first order phase transition. } for $2n_0 \lesssim n_B \lesssim 5n_0$. Each domain has own characteristic constraint; nuclear physics constraint for $n_B\lesssim 2n_0$; the $2M_\odot$ constraint for $n_B \gtrsim 5n_0$; and the causality and thermodynamic constraints on $2n_0 \lesssim n_B \lesssim 5n_0$ which require $0 \le c_s^2 \le 1$. To express the nuclear constraints for $n_B \lesssim 2n_0$, in  \cite{Kojo:2014rca} we used the Akmar-Phandharipande-Ravenhall equation of state \cite{Akmal:1998cf} as a representative. Three descriptions are matched smoothly at the boundaries.

Our quark matter description is based on a schematic quark model of the Nambu-Jona-Lasinio (NJL) type with supplemental effective vector and color-magnetic interactions. The additional interactions are included to mimic important aspects of nuclear physics and hadron spectra, and this is consistent with the crossover picture at intermediate density. In fact these interactions have played very important roles in constructing physically sensible equations of state in our 3-window treatment \cite{Kojo:2014rca}. The parameters should be regarded as those at $n_B \gtrsim 5n_0$; in general effective couplings can be medium dependent. The running vector coupling was manifestly taken into account in Ref. \cite{Fukushima:2015bda} as an illustration. In the previous studies we found that those couplings should be as large as the vacuum coupling constants, otherwise quark model equations of state easily show strong first order phase transitions to become the ideal gas equation of state, which is too soft.

We construct NG modes in this strong coupling setup. In the present model quark matter at $n_B \gtrsim 5n_0$ appears to be the color-flavor-locked (CFL) phase \cite{Alford:2007xm}. The studies of NG modes in the CFL phase are not quite new; in fact there are several elegant studies based on weak coupling pictures and effective Lagrangian approach which provided us with analytic insights \cite{Casalbuoni:1999wu,Son:1999cm,Rho:1999xf,Beane:2000ms,Bedaque:2001je}. Model studies have been done in \cite{Forbes:2004ww,Kleinhaus:2007ve,Ebert:2007bp}. 

Nevertheless we feel it necessary to update model analyses for several reasons: (i) for the construction of thermal equations of state, we need to discuss quantitative aspects of NG modes in a consistent way with the neutron equations of state at zero temperature; (ii) the strong couplings in our setup can alter weak coupling predictions at qualitative level. In particular the strong pairing effects overcome the stress in mismatched Fermi surfaces, erasing several instabilities; (iii) in the previous model studies at strong coupling, the coexistence of chiral and diquark condensates is not realized because they are separated by the first order phase transition. But if there exists a mechanism tempering the growth of quark number density, we readily find the coexistence region \cite{Kitazawa:2002bc}. In our modeling with vector interactions, the considerable amount of chiral condensates can remain at $n_B \gtrsim 5n_0$. So it is important to clarify how the residual chiral condensates affect the spectra of NG modes; (iv) the NG modes in the coexistence of chiral and diquark condensates are discussed by effective Lagrangian approach in \cite{Yamamoto:2007ah}. But this study is limited to the 3-flavor limit, and we need to add charge neutrality constraints and explicit flavor symmetry breaking for the application to neutron star physics; (v) in our strong coupling picture the gluons remain non-perturbative to $n_B \sim 10n_0$, so we should also keep the $U_A(1)$ breaking for consistency. The quark matter with non-perturbative gluons has been addressed in the context of quarkyonic matter \cite{McLerran:2007qj}.

In this work we will study the NG modes which acquire masses and effective chemical potentials through the explicit flavor symmetry breaking. They are NG modes associated with the breaking of approximate chiral symmetry and color symmetry, $SU(3)_L \times SU(3)_R \times SU(3)_c \rightarrow SU(3)_{c+L+R}$. If the $U_A(1)$ breaking by quantum effects appears to be negligible, there is an additional NG mode. As a whole we have 8+1 NG modes which can be labelled by the same quantum numbers as in the vacuum, $(\pi,K,\eta,\eta')$. We will study these modes within the random phase approximation (RPA). In this study we will not discuss the NG mode associated with the $U(1)_B$ breaking which is known to be strictly massless.

\section{A schematic quark model}
\label{sec:model}

We introduce our model Lagrangian for a 3-flavor quark matter. Our notation is as follows: the metric is $g_{\mu \nu} = {\rm diag.}(1,-1,-1,-1)$; the flavor matrices consist of $\tau_0 = \sqrt{{2/\Nf} }$ and the Gell-Mann matrices $\tau_a (a=1,\cdots, 8)$; the color matrices are the Gell-Mann matrices $\lambda_a (a=1,\cdots, 8)$; the charge matrix is $\hat{Q} = {\rm diag.}(-2/3, 1/3,1/3)$.

Our model Lagrangian for quarks and leptons, including the chemical potentials, is
\begin{equation}
\calL = \bar{q} \left( \rmi \Slash{\partial} - \hat{m} +\hat{\mu} \gamma_0 \right) q
+ \sum_{i=e,\mu} \bar{l}_i \left(\rmi \Slash{\partial}-m_i - \mu_Q \gamma_0 \right)l_i  
+ \calL_{ {\rm NJL} } + \calL_{ {\rm mag} } +  \calL_{ {\rm BB} } \,,
\end{equation}
where $\hat{m} = {\rm diag.}(m_u, m_d, m_s)$, $m_e$ and $m_\mu$ are masses for electrons and muons, and the chemical potentials for quarks are
\begin{equation}
\hat{\mu} = \mu_q + \mu^c_3 \lambda_3 + \mu^c_8 \lambda_8 + \mu_Q \hat{Q} \,.
\end{equation}
The thermodynamic variable is $\mu_q$ while the electric charge chemical potential, $\mu_Q$, and the color chemical potentials, $\mu_3^c$, $\mu_8^c$, are used as the Lagrange multipliers to guarantee the charge and color neutrality constraints, respectively.

There are several important interactions for our studies of neutron stars and the QCD phase diagram. $\calL_{ {\rm NJL} }$ is the standard NJL interaction including the chiral 4-Fermi and Kobayashi-Maskawa-'tHooft (KMT) interactions,
\begin{equation}
\calL_{ {\rm NJL} } = 
\frac{\, G_s \,}{2} \sum_{a=0}^8 \left[  (\bar{q} \lambda_a q)^2 + (\bar{q} \lambda_a \rmi \gamma_5 q)^2 \right] 
-K \left( \det_{ {\rm f} } [ \bar{q} (1-\gamma_5) q +  \det_{ {\rm f} } [ \bar{q} (1+\gamma_5) q ] \right)\,.
\end{equation}
In addition to this, we add effective interaction, $\calL_{ {\rm mag} }$, inspired from the color-magnetic interaction which play important roles to describe the pattern of the hadron spectra in quark models. It is given by
\begin{equation}
\calL_{ {\rm mag} } 
= \frac{\, H \,}{2} \sum_{A,A'=2,5,7}\left[\, 
\left(\bar{q} \, \rmi \gamma_5 \tau_A \lambda_{A'} q_C \right)
\left(\bar{q}_C \, \rmi \gamma_5 \tau_A \lambda_{A'} q \right) 
+ \left(\bar{q} \tau_A \lambda_{A'} q_C \right)
\left(\bar{q}_C  \tau_A \lambda_{A'} q \right) 
\, \right] \,,
\end{equation}
where the specific choice of the matrices $A,A'=2,5,7$ makes the attractive s-wave interactions anti-symmetric in color and flavor. Finally we consider the repulsive vector interaction,
\begin{equation}
\calL_{ {\rm BB} } = 
- \frac{\, G_V \,}{2} (\bar{q} \gamma_\mu q)^2 \,,
\end{equation}
which is inspired from the $\omega$-meson exchange between nucleons.

The model will be treated within the mean field approximation. The mean fields for the chiral condensates are $\sigma_i  =  \la\overline{q}_i q_i \ra$;
quark number density $n =\sum_{i=u,d,s} \la q^\dagger_i q_i \ra $; and diquark condensates $d_i = \la q^T C \rmi \gamma_5 R_i q \ra$ where $ \left( R_u, R_d, R_s \right) \equiv \left( \tau_7 \lambda_7, \tau_5 \lambda_5, \tau_2 \lambda_2 \right) $. With these definitions, the diquark condensates $(d_u, d_d, d_s)$ correspond to $(ds, su, ud)$ quark pairings, respectively.

The mean field particle propagators contain these mean field effects; the inverse of the propagator $S(k)$ is given by 
\begin{equation}
S^{-1} (k) = 
\begin{pmatrix}
~ \Slash{k} - \hat{M} + \hat{\mu} \gamma^0 ~&~  \gamma_5 \Delta_i R_i ~\\
~ -  \gamma_5 \Delta^\ast_i R_i ~&~ \Slash{k} - \hat{M} - \hat{\mu} \gamma^0 ~
 \end{pmatrix} ,  
  \label{eq:Sinv}
\end{equation}
where the effective Dirac and Majorana masses are given by
\begin{equation}
M_i = m_i - 2 G_s \sigma_i + K |\epsilon_{ijk}| \sigma_j \sigma_k \,,~~~~~~ \Delta_i = -H d_i \,.
\end{equation}
The thermodynamic potential for the quark part now reads
\begin{align}
\Omega_q^{ {\rm bare} } (\mu, T) 
 &=  - 2 \sum^{18}_{j=1} \int^\Lambda \!\!\frac{d^3 k}{(2\pi)^3} 
\left[ \frac{ |\epsilon_j| }{2} +T \ln \left( 1 + e^{- |\epsilon_j | /T }  \right) \right] 
\nonumber \\
& ~~
+ \sum^3_{i=1} 
\left[ G_s \sigma^2_i + H  |d_i|^2 \right] - 4 K \sigma_1 \sigma_2 \sigma_3 
- \frac{G_V}{2} n^2 ,~~~~~
\end{align}
where the single particle contributions contain $18$ independent eigenvalues, and $\Lambda$ is the UV cutoff for the Dirac sea contribution. We add the thermodynamic potential for electrons and muons, $\Omega^{ {\rm bare} }_l$, to get the full (unrenormalized) potential, $\Omega^{ {\rm bare} } = \Omega^{ {\rm bare} }_q + \Omega^{ {\rm bare} }_l$. As usual we define the renormalized thermodynamic potential as $\Omega(\mu,T) \equiv \Omega^{ {\rm bare}}(\mu,T) -  \Omega^{ {\rm bare}}(\mu=T=0)$. Then the self-consistent equations 
\begin{equation}
0 =  \frac{\, \partial \Omega \,}{\, \partial \sigma_i \,} = \frac{\, \partial \Omega \,}{\, \partial d_i \,} \,,
~~~~ n = -  \frac{\, \partial \Omega \,}{\, \partial \mu \,} \,,
\end{equation}
will be solved together with the charge neutrality conditions, $\partial \Omega/\partial \mu_{Q} =\partial \Omega/\partial \mu_{3,8}^c =0 $.

This mean field expression will be used only for $n_B \gtrsim 5n_0$, where missing confining interactions, which trap 3-quarks into a single baryon, are supposed to be less problematic than in dilute matter. Then we need to fix model parameters in medium which may differ from the vacuum values. But in the previous studies we have discussed that the $2M_\odot$ constraint disfavors in-medium coupling constants much smaller than the vacuum counterpart \cite{Kojo:2014rca,Kojo:2015fua,Fukushima:2015bda}. For this reason we keep the standard NJL parameters, $(\Lambda, m, G_s, K)$, to be at the same magnitude as the vacuum case. To be specific, we use the Hatsuda-Kunihiro parameter set \cite{Hatsuda:1994pi}:
\begin{align}
& \Lambda = 631.4\, {\rm MeV}\,,
~~~~~ G_s\Lambda^2 = 3.67\,,
~~~~~ K\Lambda^5 = 9.29 \,,
\nonumber \\
& m_{u,d} = 5.5\,{\rm MeV}\,,
~~~~~ m_s =135.7\,{\rm MeV}\,.
\end{align}
This set gives $M_{u,d}=336\,{\rm MeV}$ and $M_s=528\,{\rm MeV}$. 

The other parameters, $(g_V, H)$, are not fixed by the vacuum phenomenology for mesons, but mostly related to baryons. In our previous works, we have searched for the typical range of these parameters that can be consistent with the neutron star constraints, that is, nuclear constraints, thermodynamic and causality conditions, and the $2M_\odot$ constraint. All these constraints have played very important roles. Here we summarize our findings \cite{Kojo:2014rca,Kojo:2015fua}: (i) the value of $g_V$ should be $\gtrsim 0.5 G_s$ to keep the equations of state at $n_B \gtrsim 5n_0$ stiff enough to pass the $2M_\odot$ constraint; (ii) once $g_V$ is chosen, the constraints for $n_B \lesssim 5n_0$ can be satisfied only if the value of $H$ is tuned within $\sim 10\%$. For $g_V \gtrsim 0.5G_s$, we found that $H$ should be $\simeq 1.5G_s$. The general tendency is that as we increase $g_V$ we must also take $H$ large; there is strong correlation between allowed domains for these two parameters.

Specifically we will consider three parameter sets \cite{Kojo:2015fua}, 
\begin{equation}
(g_V, H)/G_s = (0.5, 1.4)\,, ~(0.8, 1.5)\,, ~(1.0, 1.6)~~~~~~~~{\rm for~set~(I),\, (II),\, (III)} \,.
\end{equation}
Our main results will be given for the set (II) which satisfy all constraints. In addition, we also consider the set (I) and (III) for illustrations which are slightly outside of the acceptable domain; the set (I) can satisfy the low and intermediate density constraints at $n_B \lesssim 5n_0$, but cannot make equations of state stiff enough at $n_B \gtrsim 5n_0$; meanwhile, the set (III) is stiff enough at high density but slightly violates the causality constraint. 

\begin{figure*}[!t]
\begin{center}
\hspace{-0.8cm}
\includegraphics[width = 1.03\textwidth]{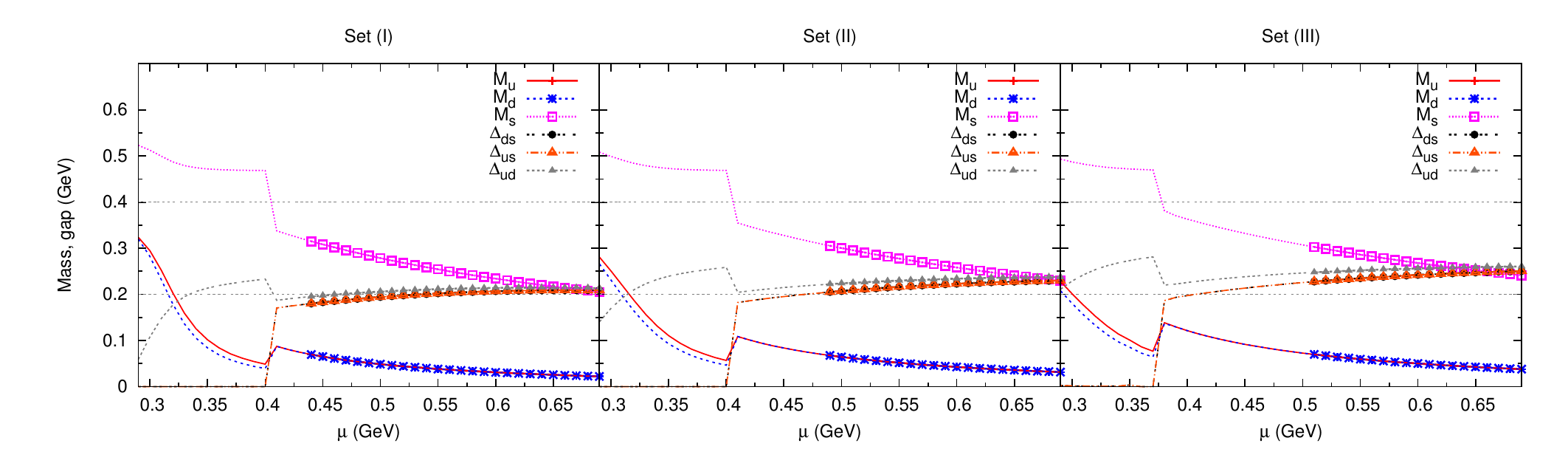}
\end{center}
\vspace{-0.2cm}
\caption{
\footnotesize{The Dirac and Majorana masses as functions of the quark chemical potential for the parameter sets (I), (II), and (III). We show the results at $n_B \gtrsim 5n_0$ by bold lines. Below $n_B < 5n_0$, the results are expected by strongly corrected by confining effects in dilute matter.
}
\vspace{-0.0cm} }\label{fig:mass_gap}
\end{figure*}

The results for Dirac and Majorana masses are shown in Fig.\ref{fig:mass_gap} for the parameter sets (I),(II), and (III). The results at $n_B < 5n_0$ are expected to be largely modified by confining effects in dilute regime. The transition between the 2-color paired phase to the CFL all occur at $n_B < 5n_0$, and it should not be taken at its face value. In the CFL region, the chiral condensates and diquark condensates coexist.

\section{Mesons in the random phase approximation}
\label{sec:RPA}
 
We study the properties of NG modes within the RPA. The method used in the present work is similar to the NJL studies in Ref. \cite{Kleinhaus:2007ve}. We extend the previous work by including the vector interaction which requires us simultaneous treatments of chiral and diquark condensates as well as anomaly terms. 

The NG modes are classified by conserved charges in the CFL phase. They are modified electric charge and modified strangeness which contain color charges in their definition. The NG modes are superposition of ordinary quark-antiquark fluctuations, $q\bar{q}$, plus diquark and anti-diquark fluctuations, $qq$ and $\bar{q} \bar{q}$, thus their properties can considerably differ from the ordinary NG modes in vacuum. Nevertheless we may still label the NG modes by $(\pi,K,\eta,\eta')$, since modified and ordinary charges appear to be the same for color-singlet objects; general NG modes including colored components can be found by adding diquark or antidiquark components, that are supplied from the diquark condensates, to ordinary color-singlet mesons with definite electric charges and strangeness.

Below we will find that neutral mesons mix considerably. For this reason, we will label neutral mesons by $(\eta_1, \eta_2, \eta_3)$ with the mass ordering $m_{\eta_1} < m_{\eta_2} < m_{\eta_3}$, instead of $(\pi_0, \eta,\eta')$. 

To study properties of NG modes, we will consider interpolating fields $\bar{q} \Gamma_{ap} q$, $q \Gamma_{pp} q$, and $\bar{q} \Gamma_{aa} \bar{q}$. A set of interpolating fields which couple one another is summarized in Table.\ref{tab:couplings}. We consider the correlators for such interpolating fields,
\begin{center}
\begin{table}[t]
\caption{\footnotesize{The interpolating fields that couple to NG modes. We list up only operators which appear in the vertex of our model Lagrangian.} }
\vspace{0.5cm}
\hspace{2.5cm}
\begin{tabular}{|c|c|c|c|c|c|}
\hline 
&~$\bar{q} \Gamma_{pa} q  $ & $\bar{q} \Gamma_{pp} q_C  $ & $\bar{q}_C \Gamma_{aa} q  $  \\ \hline 
~$\pi_- ~(\pi_+)$ ~&~$\bar{u}d $ ~($\bar{d}u $) & $ \tau_5 \lambda_7$ ($ \tau_7 \lambda_5$) & $\tau_7 \lambda_5$ ($\tau_5 \lambda_7$)\\ 
~$K_- ~(K_+)$ ~&~$\bar{u}s $ ~($\bar{s}u $) & $ \tau_2 \lambda_7$ ($ \tau_7 \lambda_2$) & $\tau_7 \lambda_2$ ($\tau_2 \lambda_7$) \\ 
~$\bar{K}_0 ~(K_0) ~$&~$\bar{d}s $ ~($\bar{s}d $) & $ \tau_2 \lambda_5$ ($ \tau_5 \lambda_2$) & $\tau_5 \lambda_2$ ($\tau_2 \lambda_5$)\\ \hline
~$\eta_1, \, \eta_2, \, \eta_3$~ &~$\bar{u}u, ~\bar{d}d, ~\bar{s}s$ ~&~ $ \tau_2 \lambda_2$, $ \tau_5 \lambda_5$, $ \tau_7 \lambda_7$ ~&~ $ \tau_2 \lambda_2$, $ \tau_5 \lambda_5$, $ \tau_7 \lambda_7$ ~ \\ \hline
\end{tabular}
\label{tab:couplings}
\end{table}
\end{center}
\begin{equation}
\rmi \Pi_{ab} (p) = \int_x \rme^{\rmi p x} \, \la 0 | J_a (x) J_b^\dag (0) | 0 \ra 
\equiv \sum_n Z_{an} Z_{nb}^* \, D_n(p_0,\vp)\,,
\end{equation}
where the operator-state coupling $Z_{an}$ and the propagator $D_n$ are given by
\begin{equation}
\la 0 | J_a (0) | n \ra = Z_{an} \,,~~~~~~
D_n(p_0,\vp) = \frac{ \rmi }{\, (p_0-\mu_n)^2 - \epsilon^2_n (\vp) \,} \,,
\end{equation}
where $\mu_n$ is the effective chemical potential arising from the charge asymmetry for the charge conjugated state we have $-\mu_n$ in place of $\mu_n$. When an effective chemical potential reaches the effective mass ($=| \epsilon(\vp =0)| $ ) of NG modes, such NG modes exhibit Bose-Einstein condensation (BEC). The kaon condensation in the CFL phase has been discussed in \cite{Bedaque:2001je,Kaplan:2001qk}.

To find out NG modes correctly within approximate methods, it is essential to keep the approximation consistent in every step. Since we have included both $(\bar{q} q)^2$ and $(qq)(\bar{q}\bar{q})$-vertices in determination of mean fields, we must also keep all of them in the RPA calculations. As a result, for a charged meson, we need to solve the RPA for $3\times 3$-coupled channel equations which include the propagation of $q\bar{q}$, $qq$, and $\bar{q}\bar{q}$. For neutral mesons, $(\pi_0, \eta, \eta')$ or $(\eta_1, \eta_2,\eta_3)$ quantum numbers may mix, so one must solve $9\times 9$-coupled channel equations.

We write its 1-loop approximation of $\Pi$ as $\Pi^0$. Then the RPA approximation leads to
\begin{equation}
\rmi \Pi
= \rmi \Pi^0 + (\rmi \Pi^0) (\rmi V)  (\rmi \Pi) 
~ \rightarrow~
\rmi \Pi =  \Pi^0 \, \frac{\rmi}{\, 1+ V \Pi^0 \,} \,
 \,,
\end{equation}
where the vertex $V$ can be read off from the pseudoscalar vertices in the Lagrangian. For the charged mesons, we read off the vertex from 
\begin{align}
\calL^{\pi_\pm} 
& = 2 \left( - G_s + K \sigma_s \right) \left( \bar{u} \gamma_5 d  \right) \left( \bar{d} \gamma_5 u  \right)
+  \frac{\, H \,}{2}  \left[\, \left(\bar{q} \tau_5 \lambda_{7} q_C \right) \left(\bar{q}_C  \tau_5 \lambda_{7} q \right) 
+ \left(\bar{q} \tau_7 \lambda_{5} q_C \right) \left(\bar{q}_C  \tau_7 \lambda_{5} q \right) \, \right]
\,, \nonumber \\
\calL^{K_\pm} 
& = 2 \left( - G_s + K \sigma_d \right) \left( \bar{u} \gamma_5 s  \right) \left( \bar{s} \gamma_5 u  \right)
+  \frac{\, H \,}{2}  \left[\, \left(\bar{q} \tau_2 \lambda_{7} q_C \right) \left(\bar{q}_C  \tau_2 \lambda_{7} q \right) 
+ \left(\bar{q} \tau_7 \lambda_{2} q_C \right) \left(\bar{q}_C  \tau_7 \lambda_{2} q \right) \, \right]
\,, \nonumber \\
\calL^{K_0, \bar{K}_0 } 
& = 2 \left( - G_s + K \sigma_u \right) \left( \bar{u} \gamma_5 d  \right) \left( \bar{d} \gamma_5 u \right)
+  \frac{\, H \,}{2}  \left[\, \left(\bar{q} \tau_2 \lambda_{5} q_C \right) \left(\bar{q}_C  \tau_2 \lambda_{5} q \right) 
+ \left(\bar{q} \tau_5 \lambda_{2} q_C \right) \left(\bar{q}_C  \tau_5 \lambda_{2} q \right) \, \right]
\,,
\end{align}
where in the KMT vertex we replace $\bar{q}q$ with the mean field value $\sigma$, following the standard recipe. For each correlator for a charged NG mode, we need $3\times 3$-matrix for the vertex $V$.

For the charge neutral sector, we have
\begin{align}
\calL_{ {\rm PS} }^{n} 
&= - G_s \left[  (\bar{u} \gamma_5 u)^2 +  (\bar{d} \gamma_5 d)^2 +  (\bar{s} \gamma_5 s)^2 \right]
+ \frac{\, H \,}{2} \sum_{A=2,5,7} \left(\bar{q} \tau_A \lambda_{A} q_C \right) \left(\bar{q}_C  \tau_A \lambda_{A} q \right) 
\nonumber \\
&~~~ 
 -2K \left[\, \sigma_s (\bar{u}\gamma_5 u) (\bar{d}\gamma_5 d)  
+ \sigma_d (\bar{u}\gamma_5 u) (\bar{s}\gamma_5 s) + \sigma_u (\bar{d}\gamma_5 d) (\bar{s}\gamma_5 s) \, \right] 
\,.
\end{align}
The resulting vertex $V$ in the RPA has the $9\times 9$ components. 

The remaining manipulations are standard. We study poles for states with $\vp=0$. We search for $p_0$ such that
\begin{equation}
1 + V \Pi_0 (p_0,\vp=0) = 0 ~~ \rightarrow ~~ p_0 = \omega_{\vp=0} \,.
\end{equation}
The $\omega_{\vp=0}$ already includes the effective chemical potential in its definition. 

The construction of mean field propagators and the RPA calculations have been done numerically. Our framework becomes those in Refs. \cite{Kleinhaus:2007ve} by dropping the terms for chiral 4-Fermi interactions and anomaly terms in the determination of condensates and the RPA. Then we have reproduced their results by choosing the same coupling constants and the UV cutoff.

In all figures, we show only the results in the CFL phase. At $n_B \lesssim 3n_0$, we have the 2SC phase which has the pairing between two colors. In this phase some quarks do not join the pairing and are gapless, so that the computations of NG modes are contaminated by the continuum, requiring the identification of complex poles. We have not performed the detailed analyses for such situations yet. In what follows, for the region $n_B \gtrsim 5n_0$ for which we apply the quark matter picture, the phase is in the CFL phase where quarks and gluons are gapped, so we will find only isolated poles. 

In the CFL region, the effective chemical potentials $\mu_Q$ and $\mu_3^c$ become almost zero, while $\mu_8^c$ is $50-100\,{\rm MeV}$. The appearance of $\mu_8^c$ causes additional splitting in the quark spectra. The lowest quark excitation appears in the channels where the $sR-uB$ or $sG-dB$ mixing occur. In these channels, two modes have the excitation energy $\sim 100\,{\rm MeV} $, smaller than the gap $\Delta_{us}, \Delta_{ds} \simeq 200\,{\rm MeV}$ because of the chemical potential effect. So the lowest possible quark-hole threshold is $\sim 200\,{\rm MeV}$, which is still above the pole of NG modes in this study.

In our strong coupling setup, we found that the mismatch among $u$-, $d$-, and $s$-quark Fermi seas are small at $n_B \gtrsim 5n_0$. As a consequence, the isospin symmetry, which is disturbed by the charge neutrality condition, is recovered in good accuracy. This leads to the degeneracy with respect to the interchange of $u$ and $d$. 

\begin{figure*}[!t]
\begin{center}
\hspace{-0.8cm}
\includegraphics[width = 1.03\textwidth]{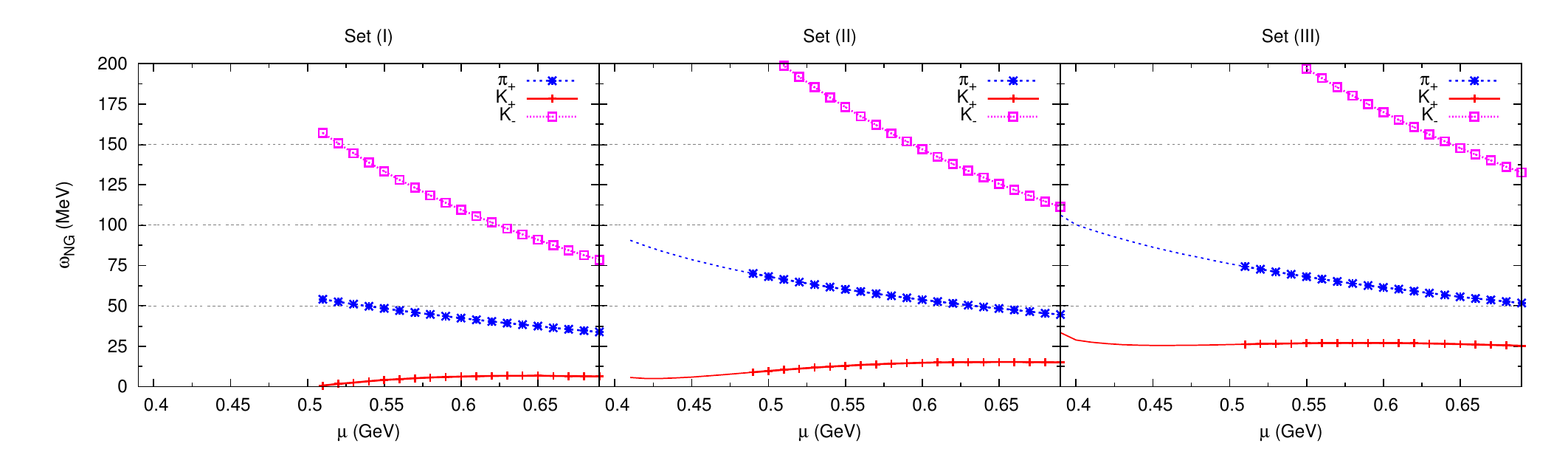}
\end{center}
\vspace{-0.2cm}
\caption{
\footnotesize{The excitation energies $\omega_{ {\rm NG} }$ for charged mesons in the parameter set (I), (II), (III). The bold lines are the results for $n_B \gtrsim 5n_0$.  In our strong coupling setup the isospin symmetry is recovered; $\omega_{\pi_+} \simeq \omega_{\pi_-}$,  $\omega_{K_+} \simeq \omega_{K_0}$, $\omega_{K_-} \simeq \omega_{K_0}$ are satisfied in very good accuracy, so we plot only $\pi_+$, $K_\pm$. For the set (I), there is the condensation of $K_+$ at $\mu_c \simeq 0.51\,{\rm GeV}$ ($n_B \simeq 6.45\, n_0$), below which our treatment is inconsistent: we must take into account the kaon condensation from the beginning.}
\vspace{-0.0cm} }\label{fig:NG_mass_c}
\end{figure*}

Shown in Fig.\ref{fig:NG_mass_c} are the excitation energies  $\omega_{ {\rm NG} }$ for charged NG modes. Because of the good isospin symmetry, the approximate relations, $\omega_{\pi_+} \simeq \omega_{\pi_-}$, $\omega_{K_+} \simeq \omega_{K_-}$, and $\omega_{K_-} \simeq \omega_{K_{0} }$, hold in good accuracy. For this reason we show only the results of $\omega_{\pi_+}$ and $\omega_{K_\pm}$. The ordering of the excitation energy is $\omega_{K_+} < \omega_{\pi_+} < \omega_{K_-}$ as suggested by previous studies. The overall excitation energies become larger as we take the values of $(g_V,H)$ larger. For the set (I), the charged kaon $K_+$ condenses at chemical potential smaller than $\mu_c \simeq 0.51\, {\rm GeV}$ ($n_B \simeq 6.45\, n_0$) as suggested by effective Lagrangian approaches\footnote{Once we find the kaon condensations, we have to go back to our mean field analyses and must include such effects from the beginning. The comprehensive studies of various CFL-K phases can be found in Ref.\cite{Warringa:2006dk}.}. However, as we increase the values of $(g_V, H)$ as required by neutron constraints, the $K_+$ becomes more massive excitations and they no longer condense for $n_B \gtrsim 5n_0$. For the set (II), typical excitation energies are $\omega_{K_+} = 5-15\, {\rm MeV}$, and for the set (III), $\omega_{K_+} = 25-30 \,{\rm MeV}$.

\begin{figure}[!t]
\begin{center}
\hspace{-0.8cm}
\includegraphics[width = 0.5\textwidth]{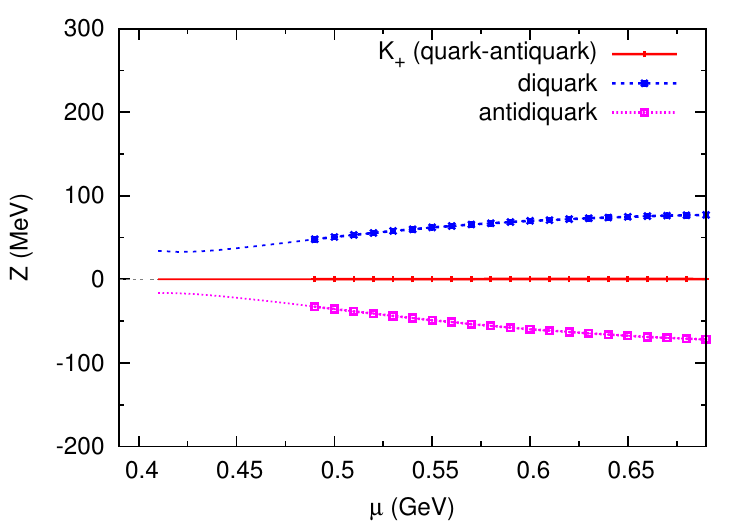}
\end{center}
\vspace{-0.2cm}
\caption{
\footnotesize{The operator-state coupling $Z$ for the $K_+$ state for the set (II). The state largely couple to diquark and antidiquark operators, while almost decouples from the quark-antiquark operators.
}
\vspace{-0.0cm} }\label{fig:resi_K_+}
\end{figure}

\begin{figure}[!t]
\begin{center}
\hspace{-0.8cm}
\includegraphics[width = 0.5\textwidth]{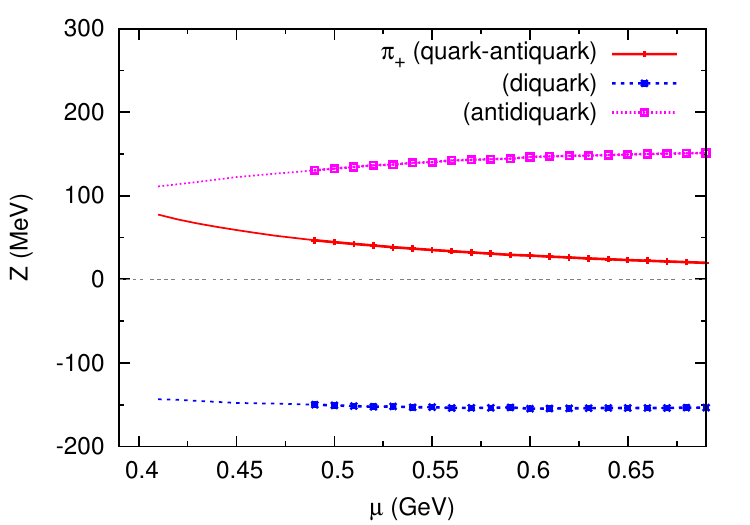}
\end{center}
\vspace{-0.2cm}
\caption{
\footnotesize{The operator-state coupling $Z$ for the $\pi_+$ state for the set (II). The state largely couple to diquark and antidiquark operators, but it also has the sizable coupling to the quark-antiquark operators which dies out as the density increases.
}
\vspace{-0.0cm} }\label{fig:resi_pi_+}
\end{figure}

From the studies of the operator-state coupling $Z_{an}$ (Fig.\ref{fig:resi_K_+}), we can clearly see that $K_+$ and $K_0$ states strongly couple to $qq$ and $\bar{q}\bar{q}$ operators, while almost decouple from $\bar{q}q$ operators. This reflects that these NG modes are dominated by modes near the Fermi surface. In contrast, the other modes such as $\pi_+$ have finite coupling strengths with $\bar{q}q$ operators (Fig.\ref{fig:resi_pi_+}), so through residual chiral condensates they carry over some features of the vacuum NG modes; these modes have larger excitation energies  and larger coupling to $\bar{q} q$ states as we approach the low density domain.

\begin{figure*}[!t]
\begin{center}
\hspace{-0.8cm}
\includegraphics[width = 1.03\textwidth]{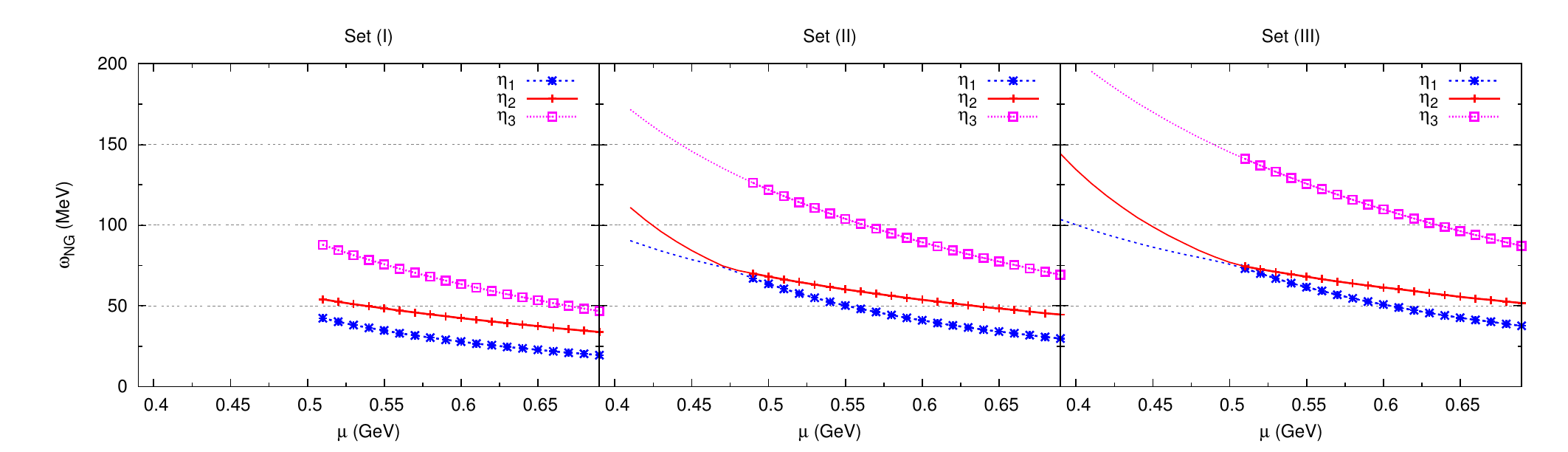}
\end{center}
\vspace{-0.2cm}
\caption{
\footnotesize{The excitation energies $\omega_{ {\rm NG} }$ for neutral mesons in the parameter set (I), (II), (III). In our strong coupling setup the isospin symmetry is recovered; $\eta_1$ has the $\pi_0$ structure at low density, and after the level repulsion $\eta_2$ at high density has the $\pi_0$ structure in good accuracy.}
\vspace{-0.0cm} }\label{fig:NG_mass_n}
\end{figure*}

Shown in Fig.\ref{fig:NG_mass_n} are the excitation energies for neutral NG modes, $(\eta_1, \eta_2, \eta_3)$. For all the parameter sets, three modes appear at low energy below $200\,{\rm MeV}$. One mode has the $\pi_0$ structure which has the degenerate spectra with $\pi_\pm$. At low density the lowest mode ($\eta_1$) has the $\pi_0$ structure but at higher density the $\pi_0$-like mode becomes the second lowest mode ($\eta_2$) after the level repulsion. The other two modes are considerable mixtures of $\eta$ and $\eta'$ quantum numbers\footnote{The mixing angle is in general the energy dependent so that those two modes do not necessarily have the orthogonal structure in the flavor space.}, in spite of our setup where the $U_A(1)$ breaking is kept.

\section{Discussions: implications for thermal equations of state}
\label{sec:thermal}

To argue the impacts of NG modes, we will adopt the parameter set (II) for which all neutron star constraints are fulfilled \cite{Kojo:2014rca}. In that setup, charged kaons do not condense, and the other mesons have masses of $30-200\,{\rm MeV}$, which are all below the continuum threshold of particle-hole excitations with gaps. The gluons should also have the gaps. Therefore our quark matter setup, all excitations, except the NG mode for the $U(1)_B$ breaking, are gapped at $n_B \gtrsim 5n_0$. 

The thermal equations of state relevant for NS-NS mergers and supernova require equations of state of $T=10-100\, {\rm MeV}$ and various lepton fractions. The details of thermal equations of state should be sensitive to the gaps for NG modes which appear to be the order of relevant temperature scale.

The thermal equations of state with gapped excitations should be contrasted with quark matter equations of state without gaps. In Ref. \cite{Masuda:2015wva}, the authors constructed thermal equations of state based on the crossover picture in a way consistent with the neutron star constraints at zero temperature. Another important work is based on the perturbative calculations for cool quark matter \cite{Kurkela:2016was}. Their constructions have gapless quarks and their thermal contributions are large, $\Delta \Omega \sim p_F^2 T^2$ ($P_F$: Fermi momentum, $T$: temperature), because of the large phase space available near the Fermi surface. In contrast, in our modeling quarks have gaps of $\sim 200\, {\rm MeV}$ near the Fermi surface and their thermal contributions should be much smaller. Then the relevant contributions start at the $\sim T^4$ or even smaller contributions (because of gaps of NG modes) from NG modes which grow slowly as $T$ increases.

For the determination of the QCD phase structure, it should be very important to distinguish gapped and gapless quark equations of state by the observations. These differences can be studied through the patterns of the gravitational waves from NS-NS mergers \cite{Read:2013zra}, if the matter reaches density as high as $n_B = 5-10 n_0$ during a merging event. Currently the gravitational waves of merger events have been discovered only for black-holes, but we may expect the detection of the gravitation waves from NS-NS mergers in near future. 

The importance of the thermal effects have been also addressed by comparing isentropic (without thermal contributions) and ideal gas equations of state for nuclear matter \cite{Bauswein:2010dn}. From our zero temperature equations of state and the gapped structure for our matter, we expect that our predictions on gravitational waves should be similar to those predicted by isentropic equations of state for nuclear matter, rather than the ideal gas types.

In summary, we study NG modes at $n_B \gtrsim 5n_0$ in a schematic quark model consistent with the zero temperature neutron star constraints. The coexistence of chiral and diquark condensates, possible residual $U_A(1)$ breaking, and explicit flavor symmetry breaking effects are taken into account. It is important to confront our results for gapped matter with the future gravitational wave detections for NS-NS mergers, since it may offer the direct test of our understanding of the QCD phase structure. For more definite predictions for gravitational wave signals, it is necessary to perform explicit construction of thermal equations of state by building a NG boson gas on our gapped quark matter. That thermal equations of state should be connected to nuclear equations of state at finite temperature \cite{Lattimer:2015nhk,Lattimer:1991nc,Shen:1998gq} via the 3-window construction. All these issues will be addressed elsewhere.

\section*{Acknowledgement}

The author acknowledges Prof. Gordon Baym for suggesting him to construct thermal equations of state. He also thanks Prof. Kenji Fukushima for several comments on NG modes in the CFL phase, and Prof. Michael Buballa for mentioning earlier important works.

\section*{References}


\end{document}